\documentclass[aps,groupedaddres8s,fleqn,nofootinbib,twocolumn]{revtex4}
\usepackage{graphicx}
\usepackage{xcolor}
\usepackage{amsmath,amssymb}
\usepackage{float}
\usepackage{physics}
\usepackage{amsfonts}
\usepackage{verbatim}
\usepackage{flushend}
\usepackage{balance}
\usepackage{endnotes}
\usepackage{footnote}
\usepackage{adjustbox}
\usepackage{gensymb}
\usepackage{subfigure}
\usepackage{float}
\usepackage{epigraph}
\usepackage{xcolor}
\usepackage[utf8]{inputenc}
\usepackage{setspace}
\newcommand{\beq}{\begin{eqnarray}}
\newcommand{\eeq}{\end{eqnarray}}
\usepackage{amsmath}

\usepackage[normalem]{ulem}

\begin{document}

\title{Viscosity bounds in liquids with different structure and bonding types}

\author{M. Withington$^{1}$, H. L. Devereux$^{1}$, C. Cockrell$^{2}$,  A. M. Elena$^{3}$, I. T. Todorov$^{3}$, Z. K. Liu${^4}$, S. L. Shang${^4}$, J. S. McCloy$^{5}$, P. A. Bingham${^6}$, K. Trachenko$^{1}$}
\address{$^1$ School of Physical and Chemical Sciences, Queen Mary University of London, Mile End Road, London, E1 4NS, UK}
\address{$^2$ Department of Materials, Imperial College London, Exhibition Road, London, SW7 2AZ, UK}
\address{$^3$ Scientific Computing Department, Science and Technology Facilities Council, Daresbury Laboratory, Keckwick Lane, Daresbury, WA4 4AD, UK}
\address{$^4$ Department of Materials Science and Engineering, The Pennsylvania State University, University Park, PA 16802, USA}
\address{$^5$ School of Mechanical and Materials Engineering, Washington State University, Pullman, WA, USA}
\address{$^6$ Materials and Engineering Research Institute, Sheffield Hallam University, Sheffield, S1 1WB, UK}

\begin{abstract}
Recently, it was realised that liquid viscosity has a lower bound which is nearly constant for all liquids and is governed by fundamental physical constants. This was supported by experimental data in noble and molecular liquids. Here, we perform large-scale molecular dynamics simulations to ascertain this bound in two other important liquid types: the ionic molten salt system LiF and metallic Pb. We find that these ionic and metallic systems similarly have lower viscosity bounds corresponding to the minimum of kinematic viscosity of about 10$^{-7}$ $\frac{{\rm m}^2}{\rm s}$. We show that this agrees with experimental data in other systems with different structures and bonding types, including noble, molecular, metallic and covalent liquids. This expands the universality of viscosity bounds into the main system types known.
\end{abstract}

\maketitle

\section{Introduction}

Different liquids are considered for use as coolants in nuclear reactors including molten salts and metals \cite{Rosenthal1970,FRANDSEN2020,Allen2007,Tang2015}. In these applications, viscosity $\eta$ and thermal conductivity are two important properties characterising the performance of these liquids and governing their flow/diffusion and thermal transports properties. Understanding and predicting these properties over a range of temperatures and pressures is therefore important from the application point of view. This understanding is also of fundamental theoretical importance, in view that properties of liquids are strongly system-dependent and hence are considered to be not amenable to a general theory, in contrast to solids and gases \cite{landaustat}.

Viscosity of low-temperature dense liquids is governed by the activation energy barrier $U$ as:

\begin{equation}
\eta=\eta_0\exp\left(\frac{U}{T}\right)
\label{eta1}
\end{equation}

\noindent where $U$ is set by the inter-molecular interactions and structure in the liquid, $T$ is temperature and $\eta_0$ is the pre-factor related to the high-temperature limit of $\eta$. Here and below, $k_{\rm B}=1$.

Eq. \eqref{eta1} applies in the low-temperature liquid-like regime where a molecule oscillates before undergoing a jump into a neighbouring quasi-equilibrium position \cite{frenkel,dyre}. In this regime, $\eta$ increases on cooling and, if crystallization is avoided, can vary by up to 15 orders magnitude in the viscous regime. At high temperature, the oscillatory component of particle motion is lost. This can occur in either the gas phase at pressures below the critical pressure or above the Frenkel line in the supercritical regime where particle dynamics become purely diffusive \cite{flreview}. In the gas-like regime of particle dynamics, $\eta$ follows a different temperature dependence:

\begin{equation}
\eta=\frac{1}{3}\rho vL
\label{eta2}
\end{equation}

\noindent where $\rho$ is density, $v$ is the average particle speed and $L$ is the particle mean free path, and where $\eta$ increases with temperature as $\eta\propto T^{\frac{1}{2}}$ because $L\propto\frac{1}{\rho}$ and $v\propto T^{\frac{1}{2}}$.

The increase of $\eta$ at low temperature and its increase at high temperature implies that $\eta$ has a {\it minimum}. It turns out that the value of the minimum of the kinematic viscosity $\nu=\frac{\eta}{\rho}$, $\nu_m$, can be approximately evaluated in terms of fundamental physical constants as \cite{sciadv1,myreview}:

\begin{equation}
v_m=\frac{1}{4\pi}\frac{\hbar}{\sqrt{m_e m}}
\label{etamin}
\end{equation}

\noindent where $m$ is the mass of the molecule and $m_e$ is the electron mass.

Deriving Eq. \eqref{etamin} involves two steps. First, it was shown that the minimum of $\nu$ depends only on two parameters characterising a condensed matter system: interatomic separation $a$ and the largest, Debye, vibration frequency $\omega_{\rm D}$. Second, characteristic values of $a$ and $\omega_{\rm D}$ are set by the Bohr radius and Rydberg energy. This gives Eq. \eqref{etamin} \cite{sciadv1,myreview}.

Eq. \eqref{etamin} has explained the long-standing question considered by Purcell and Weisskopf, namely why the viscosity of all liquids never falls below a certain value comparable to the viscosity of water at room conditions? \cite{purcell}. The answer comes in two parts \cite{pt2021}. First, viscosities stop decreasing because they have minima. Second, those minima are largely fixed by fundamental constants in Eq. \eqref{etamin} ($\nu_m\propto\frac{1}{\sqrt{m}}$ does not change $\nu_m$ too much for most liquids).

In addition to liquids, viscosity minima have been of interest in other areas of physics, including holographic models based on the correspondence between strongly-interacting field and gravity theories \cite{kss}. More generally, understanding the origin of bounds on system properties has enthralled physicists, including those interested in collective dynamics and systems where many interacting agents operate. Apart from the interest in the values and origins of the bounds themselves, there is another reason why these bounds are important: finding and understanding bounds on physical properties often means that we enhance our grasp of, or clarify the underlying physics of, the property in question \cite{grozdanov,nussinov,myreview}.

In addition to viscosity, it was realised that Eq. \eqref{etamin} provides a lower bound for an unrelated property of liquids: thermal diffusivity $\alpha$ \cite{prbthermal}. As discussed earlier, this property is directly relevant to industrial properties where molten salts are used including the operation of coolants in nuclear reactors where the heat transfer processes are important.

The lower bound $\nu_m$ in Eq. \eqref{etamin} is of the order of 10$^{-7}$ $\frac{{\rm m}^2}{\rm s}$ (this value corresponds to $m$ equal to the proton mass which sets the magnitude of atomic masses). This agrees with experimental viscosity minima for noble, molecular and network liquids to within a factor of 1-3 \cite{sciadv1}. It was also found to agree with the experimental high-temperature limiting value of viscosity of metallic alloys \cite{nussinov,nussinov2}. This analysis was further extended in Ref. \cite{khrapak}.

However, no estimations of viscosity minima in molten salts have been undertaken experimentally due to high melting points and hence very high temperatures required to reach the minima. It therefore remains unknown how the viscosities of molten salts conform to the presumably universal crossover between the liquid-like and gas-like regimes and to the theoretical minimum \eqref{etamin}. Apart from theoretical importance, knowledge of this would be important from the application point of view: knowing the pressure and temperature conditions of the minima of kinematic viscosity and thermal diffusivity would enable predictions of the optimal state of operation of molten salts.

Here, we perform large-scale molecular dynamics (MD) simulations of the molten salt LiF as a case study (LiF is a common component in molten salt mixtures used in nuclear reactors \cite{FRANDSEN2020}). We also simulate metallic Pb. We find that these ionic and metallic systems have lower viscosity bounds corresponding to the minimum of kinematic viscosity of about 10$^{-7}$ $\frac{{\rm m}^2}{\rm s}$. We show that this agrees with the experimental data in other systems with different structures and bonding types, including noble, molecular, metallic and covalent liquids. This expands the universality of viscosity bounds into the main types of systems known.

\section{Methods}

We use the DL\_POLY MD package \cite{Todorov2006}. For LiF, we used the empirical potential as in the previous work \cite{ribeiro,Kanian2009,Luo2016} with parameters derived earlier \cite{Sangster1976}. For Pb, we used the potential from Ref. \cite{Belashchenko2012}. We also simulated liquid Ar using the standard Lennard-Jones potential.

Unless otherwise stated we simulated for 1,000,000 time steps, with a fixed simulation time step of 0.001 ps. The system size used varied from 2000 atoms in LiF to 5120 in Pb. We have also simulated larger systems with 100,000 particles and found that the values of viscosity collected were consistent regardless of the system size. Similar behaviour of measured viscosities with system size has been found previously \cite{Cockrell2021}. We simulated a wide range of temperatures for each pressure. We first equilibrated the system at each pressure in the constant-pressure ensemble for 20,000 steps and then performed production runs for 1,000,000 steps in the constant energy and volume ensemble where the data were collected for calculating properties.

The dynamic viscosity $\eta$ was calculated using the Green-Kubo method \cite{Balucani1995,Zwanzig1965} as

\begin{equation}
    \eta=\frac{V}{T}\int_0^\infty dt \langle P_{xy}(0)P_{xy}(t)\rangle,
\end{equation}

\noindent where $V$ is the volume of the system and $P_{xy}$ is the $xy$ component of the stress tensor.

Obtaining accurate statistics via the Green-Kubo method is a well known computational issue \cite{Zhang2015}, which we address by averaging statistics over 20 independent initial conditions. This has been sufficient in prior work involving viscosity calculations \cite{Cockrell2021}.

We also calculate the kinematic viscosity $\nu=\frac{\eta}{\rho}$ governing the non-equilibrium flow and other properties such as, for example, fuel atomization quality \cite{Hashmi2020}. Density $\rho$ was calculated at the same state points as $\eta$.

\section{Results and discussion}

\begin{figure}
\begin{center}
{\scalebox{0.36}{\includegraphics{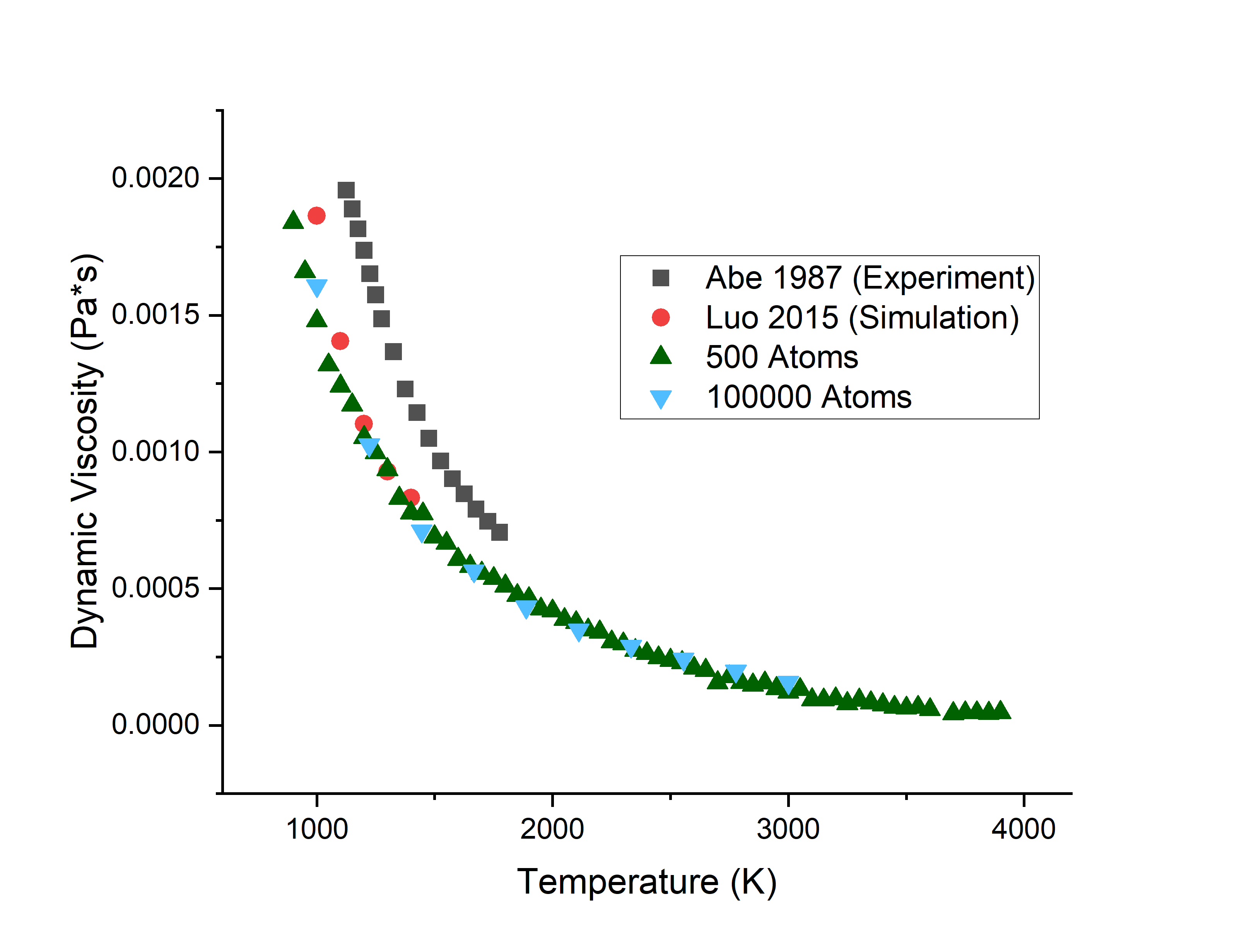}}}
{\scalebox{0.36}{\includegraphics{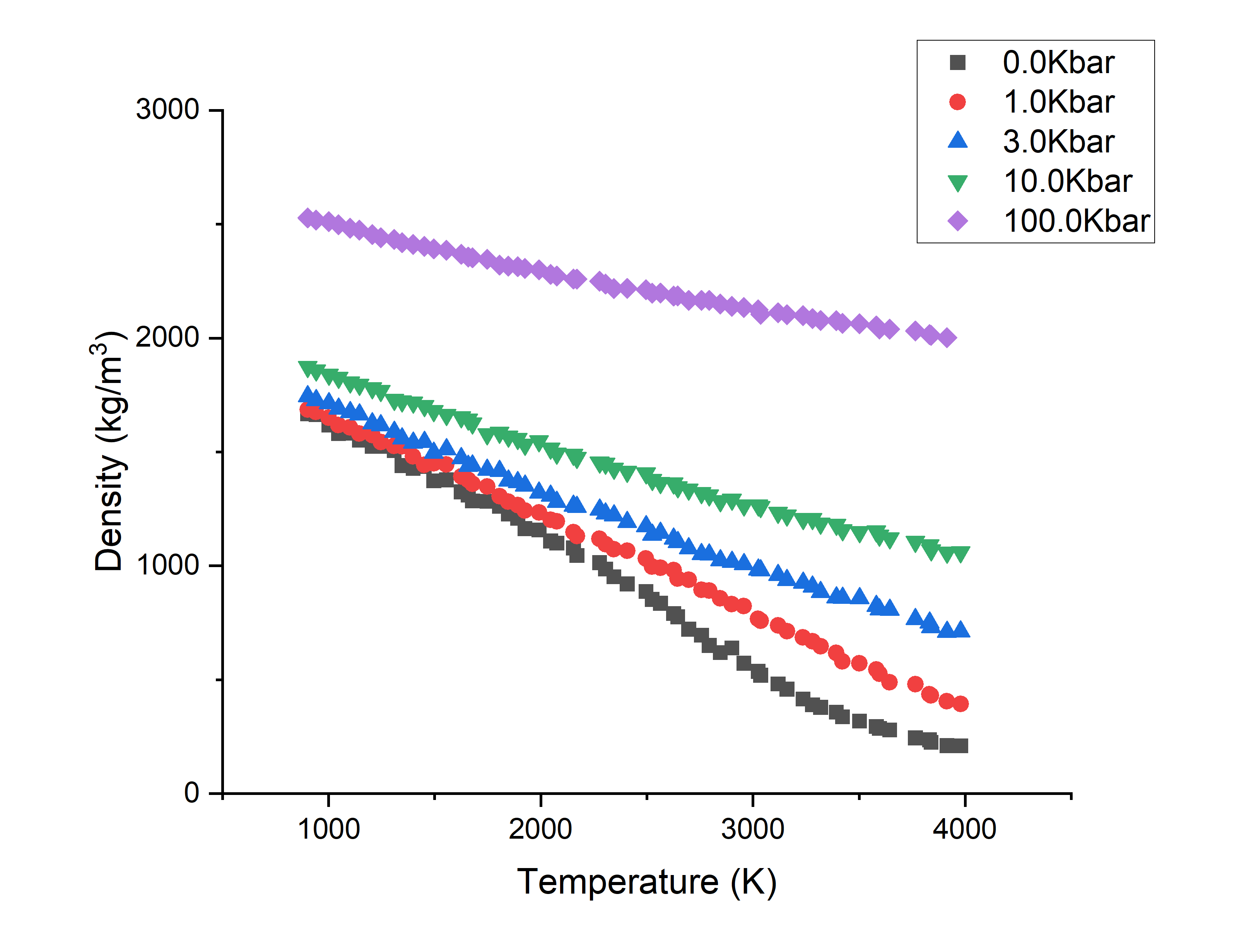}}}
{\scalebox{0.46}{\includegraphics{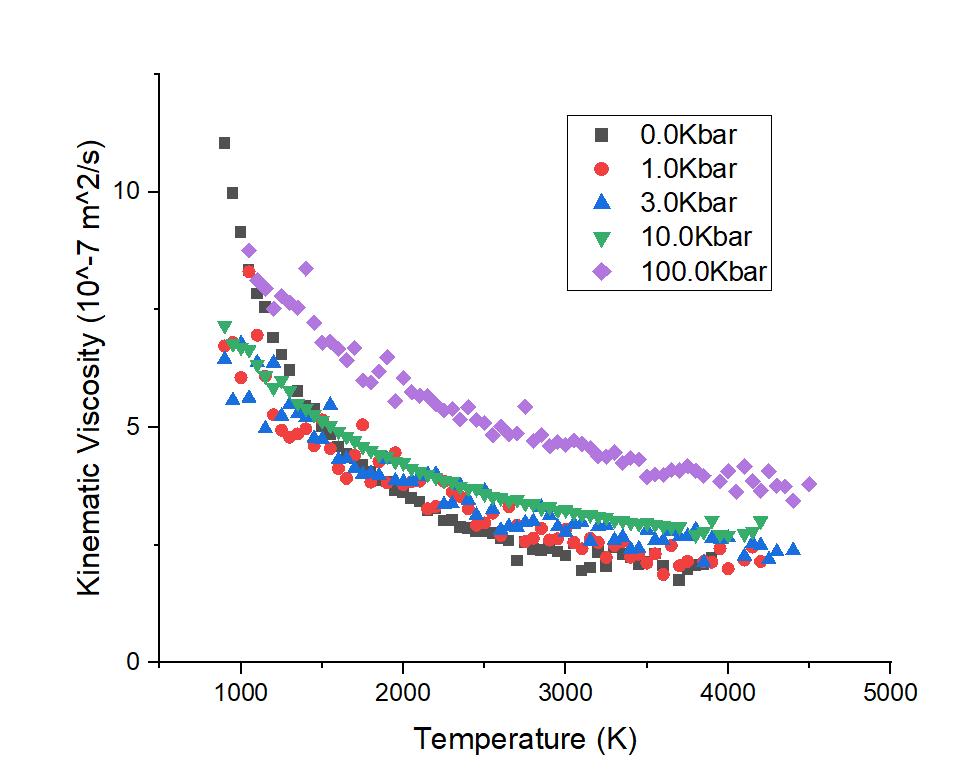}}}
\end{center}
        \caption{(a) Dynamic viscosity of LiF for small, 500 atoms, and large, 100,000 atoms, systems. We compare to earlier MD results \cite{Luo2016} as well as experimental results in \cite{Abe1981}, which has been shown to have good agreement with more recent experiments. \cite{AnaMaria2015}, \cite{Ejima1987}. (b) Density at different pressures. (c) Kinematic viscosity of LiF simulated using 2000 atoms for a range of pressures.}
        \label{visc}
\end{figure}

We show the calculated $\eta$ in Figure \ref{visc}a for two different systems sizes. We observe a good agreement of $\eta$ calculated in the two systems. We also observe the agreement with the earlier MD results in the low temperature range using the same potential \cite{Luo2016}. Little experimental data for LiF viscosity exists above 1000 K. We use the experimental data from Ref. \cite{Abe1981} which contains the widest range of data and was found to be in agreement with other experiments in the low-temperature range \cite{Merlet2010,Ejima1987}. The MD results underestimate the experimental viscosity by a factor of about 1.3-5, however, the overall shape of $\eta$ is similar in experimental and MD data. As discussed in Ref. \cite{Luo2016}, the simulated viscosity is lower than experimental viscosity due to the model not accurately matching the melting point, around 300 K lower than experimental values (predicting melting points accurately is a more general problem in atomistic simulations due to several factors including short simulation times which can be insufficient to exceed slow kinetics of phase transformations, small system sizes compared to experimental ones and hence the absence of long-wavelength fluctuations with large amplitudes which are efficient in destabilising the solid phase and so on). Translating all temperatures upwards by this melting point discrepancy of 300 K while maintaining all values of viscosity results in a much better agreement, demonstrating that the empirical potential accurately model the dissipative dynamics of the liquid. We note that the underestimation of the melting point does not affect our results since we are interested in the values of viscosity minima.

We observe that the calculated $\eta$ tends to an approximately constant value of about $\eta=2\times10^{-4}$ Pa$\cdot$s at high temperature.   This is close to viscosity minima in noble, molecular and network liquids \cite{sciadv1}. Simulating higher temperature results in the known instability of the Born-Mayer potential at short distances. This could be fixed by, for example, adding a short-range repulsive terms to the potential (e.g. in the form of Ziegler-Biersack-Littmark potentials \cite{zbl}), however, this is not required for the purposes of the current work aimed at evaluating the limiting lower viscosity bounds.

Density as a function of temperature is shown in Fig. 1b for different pressures. We observe that the slope of density decrease becomes smaller at high pressure, corresponding to the decrease of the thermal expansion coefficient with pressure. This is consistent with the common behavior seen in solids \cite{anderson1}.

The calculated kinematic viscosity $\nu=\frac{\eta}{\rho}$ is shown in Figure \ref{visc}c for different pressures. We make several observations. First, $\nu$ reaches a constant value at high temperature and shallow minima at low pressures. The minima become more apparent here, compared to in the dynamic viscosity in Figure \ref{visc}a, due to the decrease of density at high temperature. Second, pressure increases the values of the minima and constant values at which $\nu$ tends to a constant at high temperature. This is similar to what is observed in noble, molecular and network liquids \cite{sciadv1} and is related to the increase of the activation energy for molecular rearrangement with pressure and associated increase of viscosity. Third, the value of $\nu$ at their constant value or minima is in the range (2-5)$\cdot 10^{-7}\frac{{\rm m}^2}{\rm s}$. This is of the same order, $10^{-7}\frac{{\rm m}^2}{\rm s}$ as predicted by Eq. \eqref{etamin}. This last point is important and implies that the viscosity of ionic systems and molten salts have the same behavior in terms of the value of their viscosity minima. We will make the comparison between calculated and theoretical values more quantitative later on, alongside the other liquids we study.

\begin{figure}
\begin{center}
{\scalebox{0.46}{\includegraphics{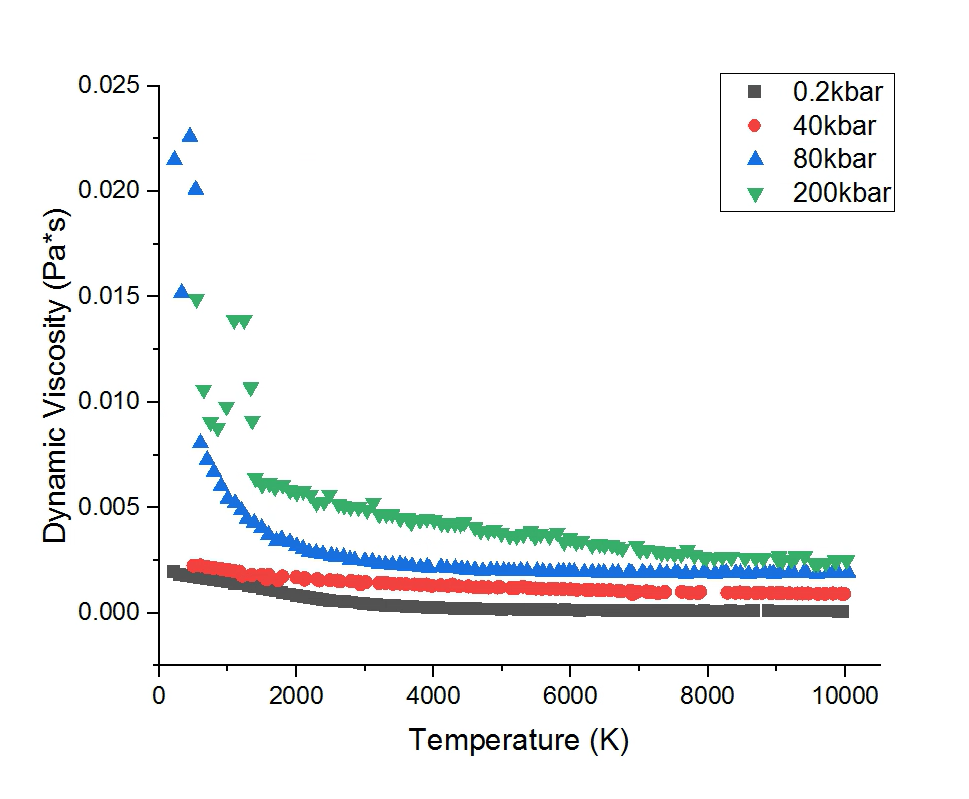}}}
{\scalebox{0.3}{\includegraphics{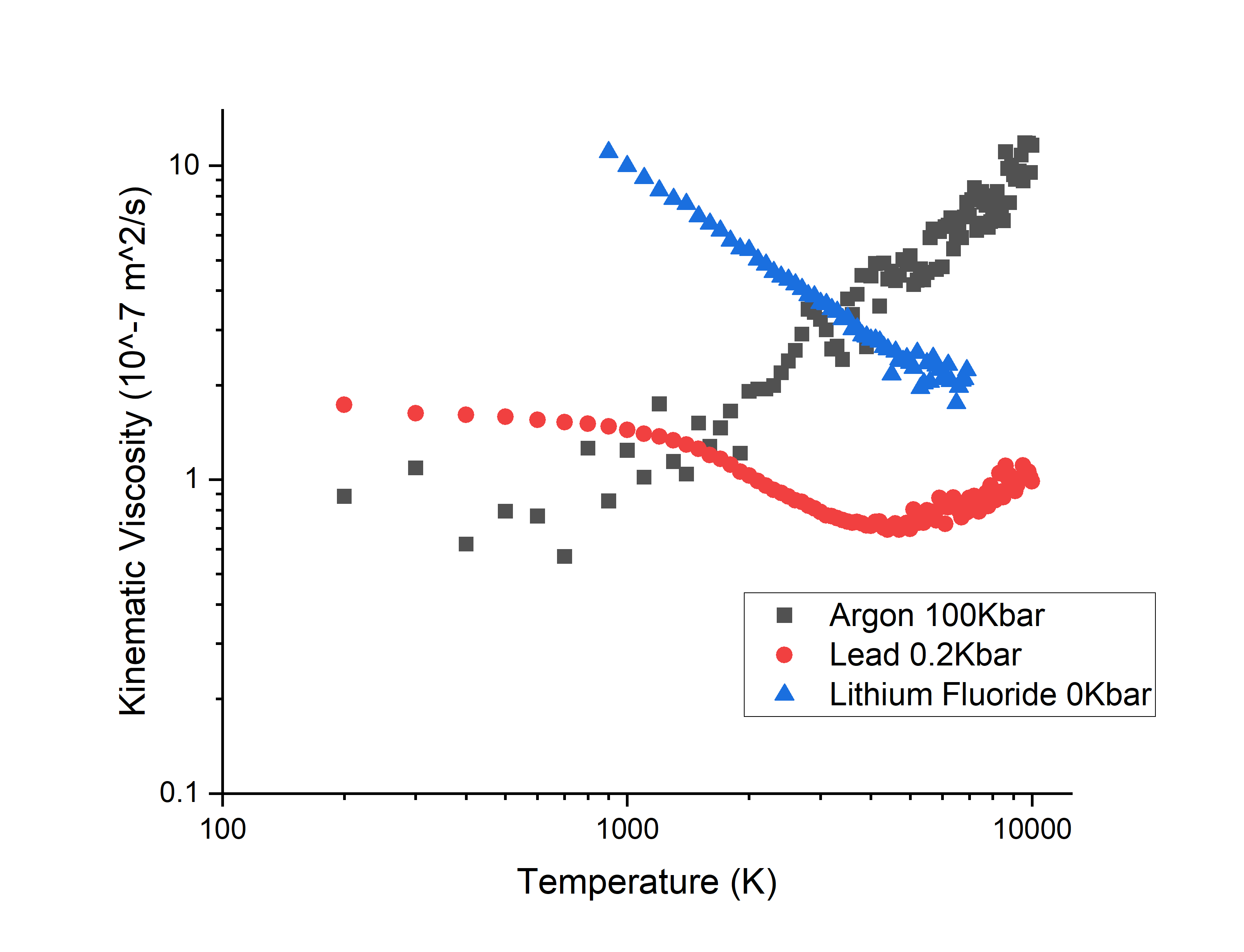}}}
\end{center}
        \caption{(a) Dynamic Viscosity of Pb at 0.2 kbar, 40 kbar, 80 kbar and 200 kbar. (b) Kinematic viscosity of Pb. (c) Comparison of kinematic viscosity of Ar, Pb and LiF.}
\label{fig:LeadDynam_kin}
\end{figure}

$\eta$ and $\nu$ for the simulated metallic system, Pb, are shown in Fig. \ref{fig:LeadDynam_kin}. These show very similar results to those of LiF. $\eta$ increases with pressure and tends to about $\eta=2\times10^{-4}$ Pa$\cdot$s at high temperature. $\nu$ has a minimum around $2\cdot 10^{-7}\frac{{\rm m}^2}{\rm s}$ at low pressure as predicted by Eq. \eqref{etamin}. To emphasise this similarity, we plot $\nu$ for Pb, LiF and Ar in Fig. \ref{fig:LeadDynam_kin}c.

\begin{figure}
\begin{center}
{\scalebox{0.36}{\includegraphics{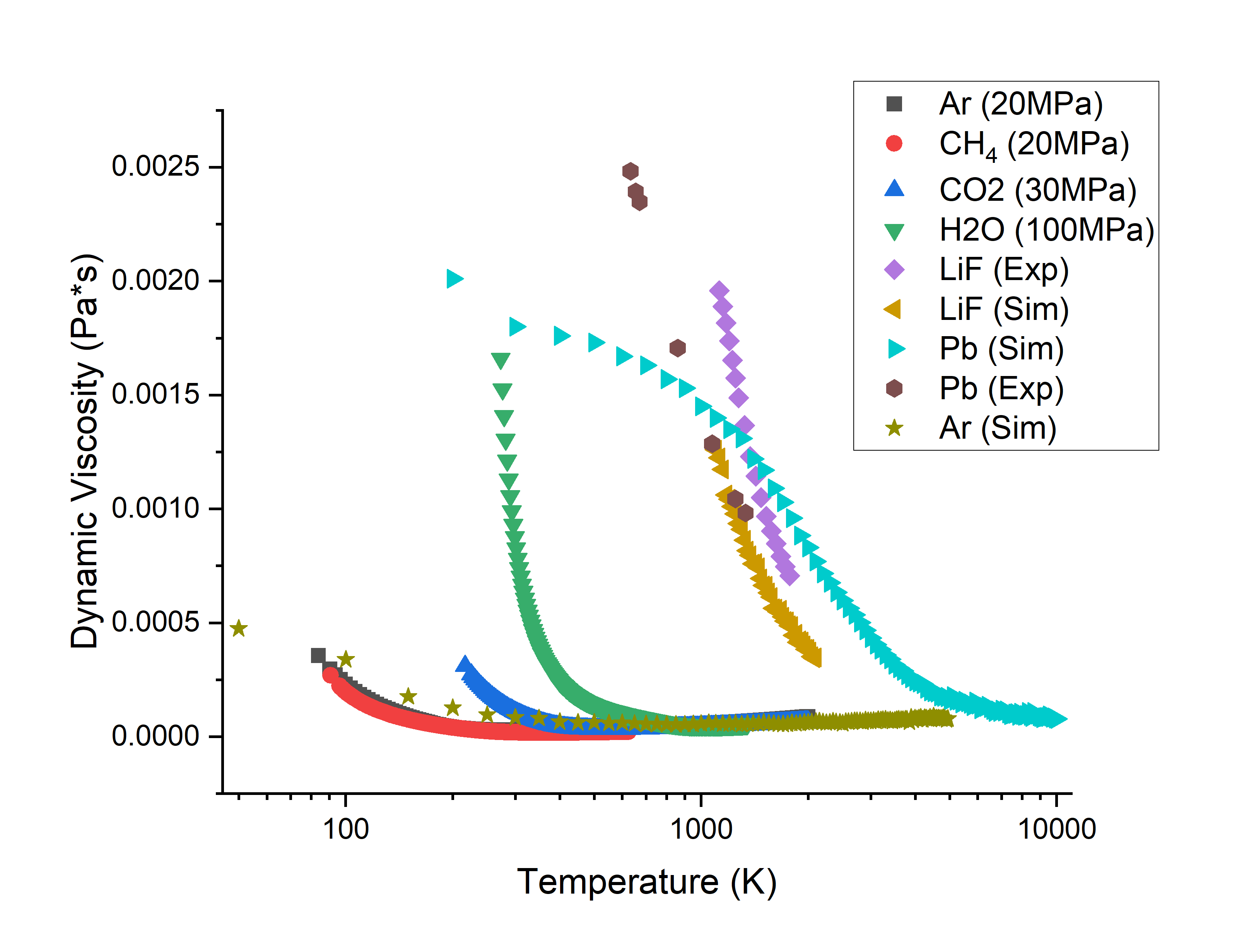}}}
{\scalebox{0.33}{\includegraphics{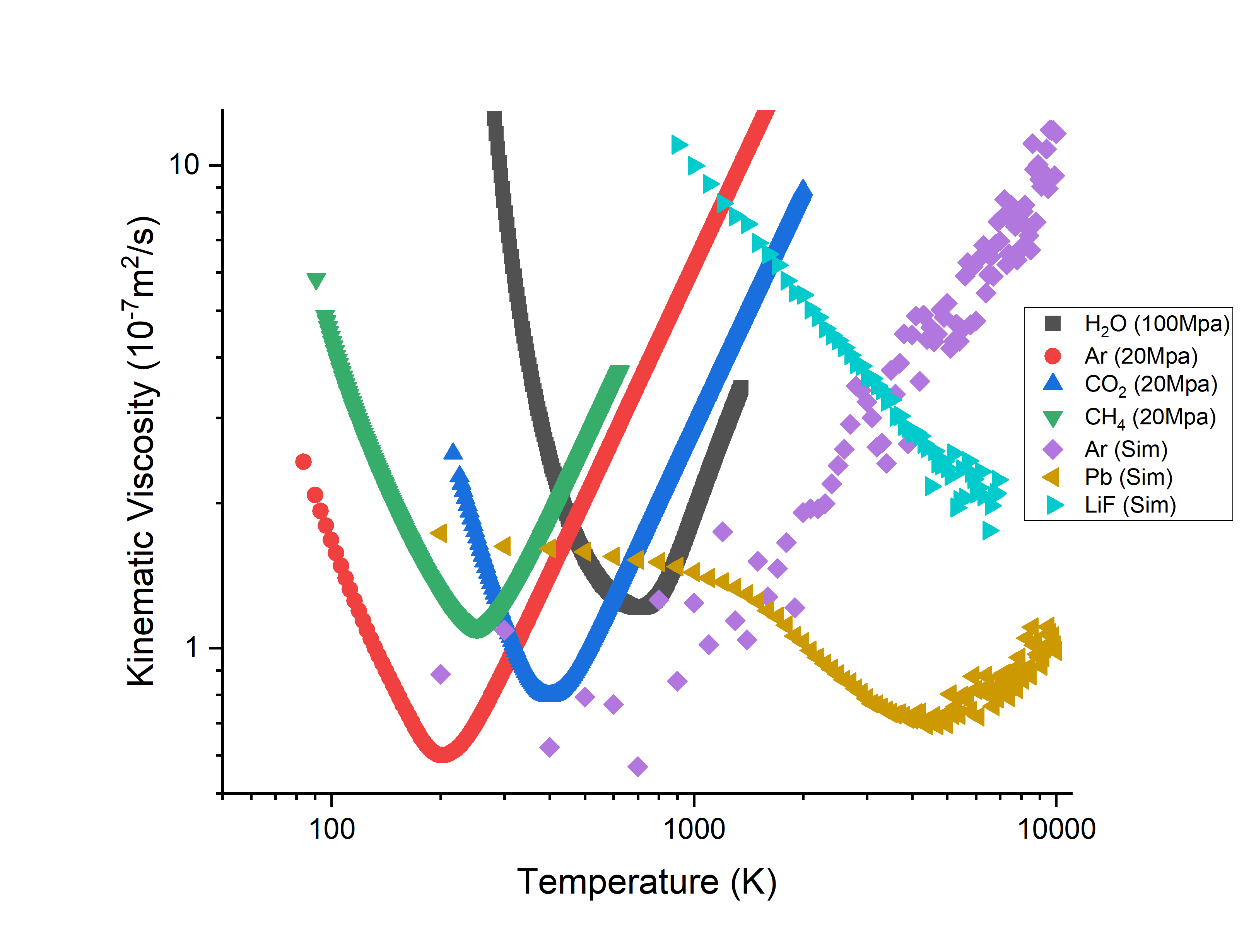}}}
\end{center}
        \caption{(a) Experimental and simulated dynamic viscosity for LiF, Pb and Ar. In order: NIST data set for Ar 20 MPa, CH$_4$ 20 MPa, CO$_2$ 30 MPa, H$_2$0 100 MPa \cite{nist}, LiF experiment \cite{Abe1981}, our LiF simulation,our Pb simulation, Pb experiment \cite{Ofte1967} and our Ar simulation. (b) Kinematic viscosity of systems in (a).}
\label{fig:dynam_exp}
\end{figure}

To compare our results for ionic LiF and metallic Pb with a wider data set, we add the experimental $\eta$ and $\nu$ for noble Ar and molecular CH$_4$ and CO$_2$ as well as network H$_2$O \cite{nist} to the plots in Fig. \ref{fig:dynam_exp}.
It is appropriate to note here that, differently from noble and molecular liquids, the NIST database \cite{nist} does not include viscosities of ionic, molten salts and metallic systems. Part of the issue is the experimental difficulty related to high melting and boiling points of these systems. The value of our current simulations therefore includes provision of this data and serving as a guide for future high-temperature experiments.

The results collected in Fig. \ref{fig:dynam_exp} are consistent with what is expected for the dependence of viscosity on temperature, as discussed in the Introduction: the decrease with temperature in the liquid-like regime, followed by its increase in the gas-like regime. We observe a great amount of variation in the values of viscosity collected for different systems, as well as the variability of the shape of viscosity curves. We also observe a consistency in the magnitude of the minimum values of viscosity, in agreement with Eq. \eqref{etamin} predicting close values of the lower viscosity bound in all liquids. This includes the ionic molten salt LiF and metallic Pb.

We now address the quantitative comparison between the predicted and calculated $\nu_m$. In Table \ref{tab:calc_vs_sim} we show theoretical and calculated $\nu_m$ for LiF, Pb and Ar. The ratio between calculated and theoretical values is in the range 1.7-4.6. This is similar to the range of ratios for the large set of noble, molecular and network liquids found earlier, where this ratio is in the range 0.5-3 \cite{sciadv1}.

\begin{table}[ht]
        \begin{tabular}{ l l l }
         \hline
         Liquid & $v_m 10^{-8}$ $\frac{{\rm m}^2}{\rm s}$ (theor.) & $v_m 10^{-8}$ $\frac{{\rm m}^2}{\rm s}$ (sim.)  \\
         \hline
         LiF &  6.0 & 18.8 \\
         Pb & 1.5 & 6.9\\
         Ar & 3.4 & 5.7 \\
         \hline
         Liquid & $v_m 10^{-8}$ $\frac{{\rm m}^2}{\rm s}$ (theor.) & $v_m 10^{-8}$ $\frac{{\rm m}^2}{\rm s}$ (exp.)  \\
         \hline
         Ar (20 MPa) & 3.4 & 5.9 \\
         CH$_4$ (20 Mpa) & 5.4 & 11.0\\
         CO$_2$ (30 Mpa) & 3.2 & 8.0\\
         H$_2$O (100 MPa)& 5.1 & 12.1\\
         \hline
    \end{tabular}
    \caption{Comparison of predicted values calculated from the formula against simulated or experimental values of the kinematic viscosity minimum.}
    \label{tab:calc_vs_sim}
\end{table}

The difference between theoretical and observed $\nu_m$ by a factor of 1-4 is related to a number of approximations involved in deriving Eq. \eqref{etamin}. This includes approximating the interatomic separation by the Bohr radius and the bonding energy by the Rydberg energy. Recall that the main purpose of Eq. \eqref{etamin} is two-fold: first, it shows that viscosity minimum is largely the same for all liquids because it is set by fundamental physical constants. Second, Eq. \eqref{etamin} serves to evaluate a characteristic value, order of magnitude, of $\nu_m$.

We have discussed viscosity minima in ionic, metallic, noble, molecular and network liquids. This leaves the remaining important bonding type: covalent. Silica is a commonly discussed system in which covalency is strong: it is a system with mixed bonding where covalency and ionicity make similar contributions to the bonding type. Experimentally, viscosity of silica was measured up to about 3000 K, and yet no minimum was seen due to challenges involved in high-temperature experiments \cite{ojovan}. Simulated silica, taken to yet higher temperature in excess of 6000 K, shows saturation to a constant value of about $10^{-3}$ Pa$\cdot$s \cite{horbach}. This corresponds to $\nu_m$ of about 5$\cdot 10^{-7}$ $\frac{{\rm m}^2}{\rm s}$ and falls in the range of $\nu_m$ predicted theoretically by Eq. \eqref{etamin}.

An interesting observation from Eq. \eqref{etamin} is that the minimal viscosity is a quantum property and approaches zero in the classical limit $\hbar=0$. This might be perceived to be at odds with our thinking about liquids as mostly high-temperature classical systems. However, the nature and origin of interatomic forces and radii are quantum-mechanical, and Eq. \eqref{etamin} reminds us of this \cite{pt2021}. This then brings the question of how our classical MD simulations reproduce the lower bound of liquid viscosity, the essentially quantum effect? We understand this if we recall that the parameters of empirical potentials in classical MD simulations are tuned to result in interatomic separations and energy values in real systems (either by fitting to experiments or quantum-mechanical simulations) where they are quantum-mechanical in origin. The classical simulations capture quantum effects, including the lower viscosity bound, through these potential parameters.

\section{Summary}

We have explored the nature of viscosity minima in ionic molten salt liquid LiF, complemented by metallic Pb. We have found that these systems have lower viscosity bounds corresponding to the minimum of kinematic viscosity of about 10$^{-7}$ $\frac{{\rm m}^2}{\rm s}$. This agrees with the experimental data for other systems with different structures and bonding type, including noble, molecular, metallic and covalent liquids, and it expands the universality of viscosity bounds into the main types of systems known. In future work, it may be interesting to develop more accurate potentials or employ quantum-mechanical molecular dynamics simulations to simulate molten salts.

We have previously found that Eq. \eqref{etamin} also give the minima of an unrelated, yet important property: thermal diffusivity \cite{prbthermal}. This was supported by the experimental data for noble and molecular systems. Our current findings therefore suggest that, similarly to the kinematic viscosity, the minima of thermal diffusivity are more universal and include other system types including the ionic molten salts. This is interesting to explore in future work.

We are grateful to V. V. Brazhkin for discussions, EPSRC (grants No. EP/X011607/1 and EP/W029006/1) and Queen Mary University of London for support. Computing resources for the $10^5$ LiF data were provided by STFC Scientific Computing Department’s SCARF cluster. For all other simulations conducted, this research utilised Queen Mary's Apocrita HPC facility, supported by QMUL Research-IT. http://doi.org/10.5281/zenodo.438045. The U.S. researchers acknowledge the financial support from the U.S. Department of Energy via Grant no. DE-NE0009288.


\end{document}